\documentclass[aps,prc,twocolumn,superscriptaddress,showkeys]{revtex4-1}

\usepackage{amsmath}
\usepackage{graphicx}
\usepackage{bm}
\usepackage{booktabs}
\usepackage{multirow}
\usepackage{revsymb}
\usepackage{threeparttable}
\usepackage{CJKutf8}
\newcommand{\cn}[1]{\begin{CJK*}{UTF8}{gbsn}#1\end{CJK*}}

\begin{document}

\title{Proton-to-Alpha Branching Ratio in the $^{12}$C+$^{12}$C fusion reaction at Astrophysical Energies}

\author{Ruojun Yang (\cn{杨若君})}
\affiliation{Sino-French Institute of Nuclear Engineering and Technology, Sun Yat-sen University, Zhuhai, Guangdong 519082, China}

\author{Ruiqi Chen (\cn{陈锐麒})}
\affiliation{Sino-French Institute of Nuclear Engineering and Technology, Sun Yat-sen University, Zhuhai, Guangdong 519082, China}

\author{Xiao Fang (\cn{方晓})}  %%%  Corresponding author
\email{fangx26@mail.sysu.edu.cn}
\affiliation{Sino-French Institute of Nuclear Engineering and Technology, Sun Yat-sen University, Zhuhai, Guangdong 519082, China}

\author{Yihua Fan (\cn{范翊华})}
\affiliation{Institute of Modern Physics, Chinese Academy of Sciences, Lanzhou, Gansu 730000, China}
\affiliation{School of Nuclear Science and Technology, University of Chinese Academy of Sciences, Beijing 100049, China}

\author{Xiaodong Tang (\cn{唐晓东})}  %%%  Corresponding author
\email{xtang@impcas.ac.cn}
\affiliation{Institute of Modern Physics, Chinese Academy of Sciences, Lanzhou, Gansu 730000, China}
\affiliation{School of Nuclear Science and Technology, University of Chinese Academy of Sciences, Beijing 100049, China}

\author{Yunju Li (\cn{李云居})}  %%%  Corresponding author
\email{li\_yunju@163.com}
\affiliation{China Institute of Atomic Energy, Beijing 102413, China}

\author{Fengqiao Luo (\cn{罗枫桥})}
\affiliation{University of Notre Dame, Notre Dame, IN 46556, USA}

\date{\today}

\begin{abstract}
The unique resonance features in the $^{12}$C+$^{12}$C fusion reaction lead to significant fluctuations in the branching ratio $R_{p/\alpha}=\sigma_p/\sigma_\alpha$, making it difficult to determine the $R_{p/\alpha}$ at astrophysical energies. By combining Hauser--Feshbach statistical-model calculations with constraints from direct charged-particle and gamma-ray measurements, we investigate the energy dependence of the averaged $R_{p/\alpha}$ and predict its behavior within the Gamow window. 
Owing to the strong energy dependence of $R_{p/\alpha}$, the corresponding reaction-rate ratios, $\langle \sigma v \rangle_p / \langle \sigma v \rangle_\alpha$, during core and shell carbon burning are determined to be 0.29, 0.45, and 0.52 at $T_9 = 0.5$, 1.0, and 1.2, respectively, significantly lower than the widely adopted CF88 constant value of 0.79. The implications of the revised $\langle \sigma v \rangle_p / \langle \sigma v \rangle_\alpha$ ratio for stellar nucleosynthesis and white-dwarf evolution are also discussed.
\end{abstract}

\keywords{carbon burning, Hauser-Feshbach, Talys, branching ratio, $R_{p/\alpha}=\sigma_{p}/\sigma_{\alpha}$}

\maketitle

\section{Introduction}

The $^{12}$C+$^{12}$C fusion reaction is a key process in stellar evolution, governing the carbon-burning phase in massive stars, the formation of white dwarfs, and the ignition of type-Ia supernovae and superbursts \cite{Gasques2005,DeGeronimo2024,chieffi2025status,Wiescher2025Quantum}. At astrophysical temperatures ($0.5 \leq T_9 \leq 1.2$), the reaction proceeds at sub-barrier energies corresponding to $E_{\mathrm{cm}} \sim 1$--$3$ MeV.

The fusion forms a highly excited $^{24}$Mg compound nucleus with excitation energies of 15--17 MeV, which predominantly decays through the proton and $\alpha$ channels, $^{12}{\rm C}(^{12}{\rm C},p){}^{23}{\rm Na}$ and $^{12}{\rm C}(^{12}{\rm C},\alpha){}^{20}{\rm Ne}$. The branching ratio is
\begin{equation}
R_{p/\alpha}(E) = \frac{\sigma_p (E)}{\sigma_\alpha (E)} = \frac{S^*_p(E)}{S^*_\alpha (E)},
\end{equation}
where $\sigma_p$ and $\sigma_\alpha$ are the fusion cross sections for the $^{12}{\rm C}(^{12}{\rm C},p){}^{23}{\rm Na}$ and
$^{12}{\rm C}(^{12}{\rm C},\alpha){}^{20}{\rm Ne}$ channels, respectively, and $S^*$ is the modified astrophysical S-factor, quantifies the competition between these two main reaction channels. $S^*$ is defined as~\cite{patterson1969}
\begin{equation}
S^{*}(E_{\rm cm}) = \sigma(E_{\rm cm}) E_{\rm cm} \exp\left(\frac{87.21}{\sqrt{E_{\rm cm}}} + 0.46 E_{\rm cm}\right),
\label{sfactor}
\end{equation}
where $\sigma(E_{\rm cm})$ is the fusion cross section and $E_{\rm cm}$ is the center-of-mass energy in units of MeV. The modified S-factors for the
$^{12}{\rm C}(^{12}{\rm C},p){}^{23}{\rm Na}$ and
$^{12}{\rm C}(^{12}{\rm C},\alpha){}^{20}{\rm Ne}$ channels are denoted by
$S^{*}_{p}$ and $S^{*}_{\alpha}$, respectively.

This ratio has direct astrophysical implications. It governs the relative production of $^{23}$Na and $^{20}$Ne and influences nucleosynthesis pathways in massive stars~\cite{pignatari2012}. It also serves as an important input for white dwarf models \cite{DeGeronimo2024}, particularly in interpreting observed stellar seismology.

The independent investigation of the proton-to-alpha branching ratio, $R_{p/\alpha} = \sigma_p / \sigma_\alpha$, is highly motivated by its distinct advantages over absolute cross-section measurements. Experimentally, extracting this ratio effectively cancels out major sources of systematic uncertainties, such as those originating from target thickness, beam current integration, and the absolute calibration of detection efficiencies. Consequently, $R_{p/\alpha}$ serves as an exceptionally clean and robust probe for exploring compound-nucleus decay dynamics.

Although numerous experiments on the $^{12}\mathrm{C}+^{12}\mathrm{C}$ fusion reaction have been performed over the past decades, significant discrepancies still remain among the reported cross sections, particularly at low astrophysical energies. In Ref.~\cite{li2020modified}, Li \textit{et al.} employed the statistical-model code TALYS~\cite{koning2005,koning2023talys} to estimate the branching ratios of unobserved (missing) channels in both the proton and $\alpha$ decay modes. By correcting the measured partial cross sections using these estimated branching ratios, they reconciled a substantial fraction of the discrepancies among the existing data sets. After applying these corrections, the experimental results became consistent within approximately $\pm 30\%$ for $E_{\mathrm{cm}} \gtrsim 4~\mathrm{MeV}$. The total fusion cross section was subsequently obtained by combining the proton, $\alpha$, and neutron channels, and was compared with the proposed upper and lower limits.

In this work, we investigate the averaged behavior of the ratio $R_{p/\alpha}(E)$ by normalizing TALYS-based calculations from Ref.~\cite{li2020modified} to experimental cross sections corrected for missing channels. 
This approach enables us to derive an updated description of the competition between the $^{12}{\rm C}(^{12}{\rm C},p){}^{23}{\rm Na}$ and $^{12}{\rm C}(^{12}{\rm C},\alpha){}^{20}{\rm Ne}$ channels. We further perform an extrapolation to astrophysical energies and provide a revised evaluation of the corresponding reaction-rate ratio. The astrophysical implications are also discussed.

\section{Theoretical Framework}

The decay is described using the Hauser-Feshbach statistical model \cite{hauser1952}. For a reaction from an entrance channel $\alpha$ to an exit channel $\alpha'$, the averaged cross-section can be written as \cite{hauser1952}
%\begin{equation}
\begin{align}
\nonumber
\sigma_{\alpha \alpha'} 
&= \pi \lambdabar_\alpha^2 \times \\
&
\sum_{J} 
\frac{2J + 1}{(2I + 1)(2i + 1)}
\,
\frac{
\left\{ \sum_{S l} T_l(\alpha) \right\}^{J}
\left\{ \sum_{S' l'} T_{l'}(\alpha') \right\}^{J}
}{
\left\{ \sum_{\alpha'', S'' l''} T_{l''}(\alpha'') \right\}^{J}
},
\label{eq:HF_original}
\end{align}
%\end{equation}
where $T_l$ denotes the optical-model transmission coefficient for a given partial wave.

A useful interpretation of Eq.~\eqref{eq:HF_original} can be obtained in terms of decay branching ratios. 
For a given compound nucleus state with total angular momentum $J$, the probability that the system decays into a specific exit channel $\alpha'$ is determined by the relative strength of its transmission coefficient compared to all energetically accessible channels. This defines the branching ratio
\begin{equation}
b_{\alpha'}^J = \frac{
\sum_{S' l'} T_{l'}(\alpha') ^{J}
}{
 \sum_{\alpha'', S'' l''} T_{l''}(\alpha'') ^{J}
},
\end{equation}
where the denominator represents the total transmission strength over all open channels. 

Furthermore, it is important to note that the entrance channel consists of two identical $^{12}\text{C}$ bosons with $J^{\pi} = 0^+$. Due to the requirement of totally symmetric spatial wave functions, only even partial waves ($L = 0, 2, 4, \dots$) are permitted in the entrance channel. This parity-dependent selection rule explicitly modifies the entrance-channel transmission coefficients $T_l$ and is rigorously considered in our statistical calculations. The spin populations of the $^{24}\text{Mg}$ compound nucleus produced by $^{12}$C+$^{12}$C fusion, calculated with CCFULL code \cite{hagino1999}, are displayed in the Fig. 4 of Ref. \cite{li2020modified}. At sub-barrier energies, the spin-parity population of the $^{24}\text{Mg}$ compound nucleus is dominated by the 0\(^+\), 2\(^+\), 4\(^+\), and 6\(^+\) components. As the energy decreases, high-spin fusion channels are rapidly suppressed and become inaccessible. Over the \(E_\text{cm} = 1\)--3 MeV stellar energy range, the 2\(^+\) state consistently accounts for approximately 56\% of the total population, while the 0\(^+\) fraction decreases monotonically from 39\% at \(E_\text{cm}=1\) MeV to 33\% at \(E_\text{cm}=3\) MeV.

The statistical model calculations were conducted via TALYS, where the reaction mechanism was constrained to the compound mode, excluding pre-equilibrium and direct reaction components to maintain consistency with the low-energy physics of the $^{12}$C+$^{12}$C system. The default global Optical Model Potential (OMP) was employed for both nucleons and $\alpha$ particles. The nuclear level density was calculated using the Constant Temperature Model matching with the Fermi Gas Model, and width fluctuation corrections were enabled. The calculations were performed separately for each individual spin state, with the final macroscopic results constructed by weighting these discrete contributions according to the spin population predicted by CCFULL. Further computational details can be found in Ref. \cite{li2020modified}.

\section{Averaged Trend of the $R_{p/\alpha}$ ratio}
\label{sec_prediction}

\begin{figure*}[!htbp]
	\centering
	\includegraphics[width=0.88\textwidth]{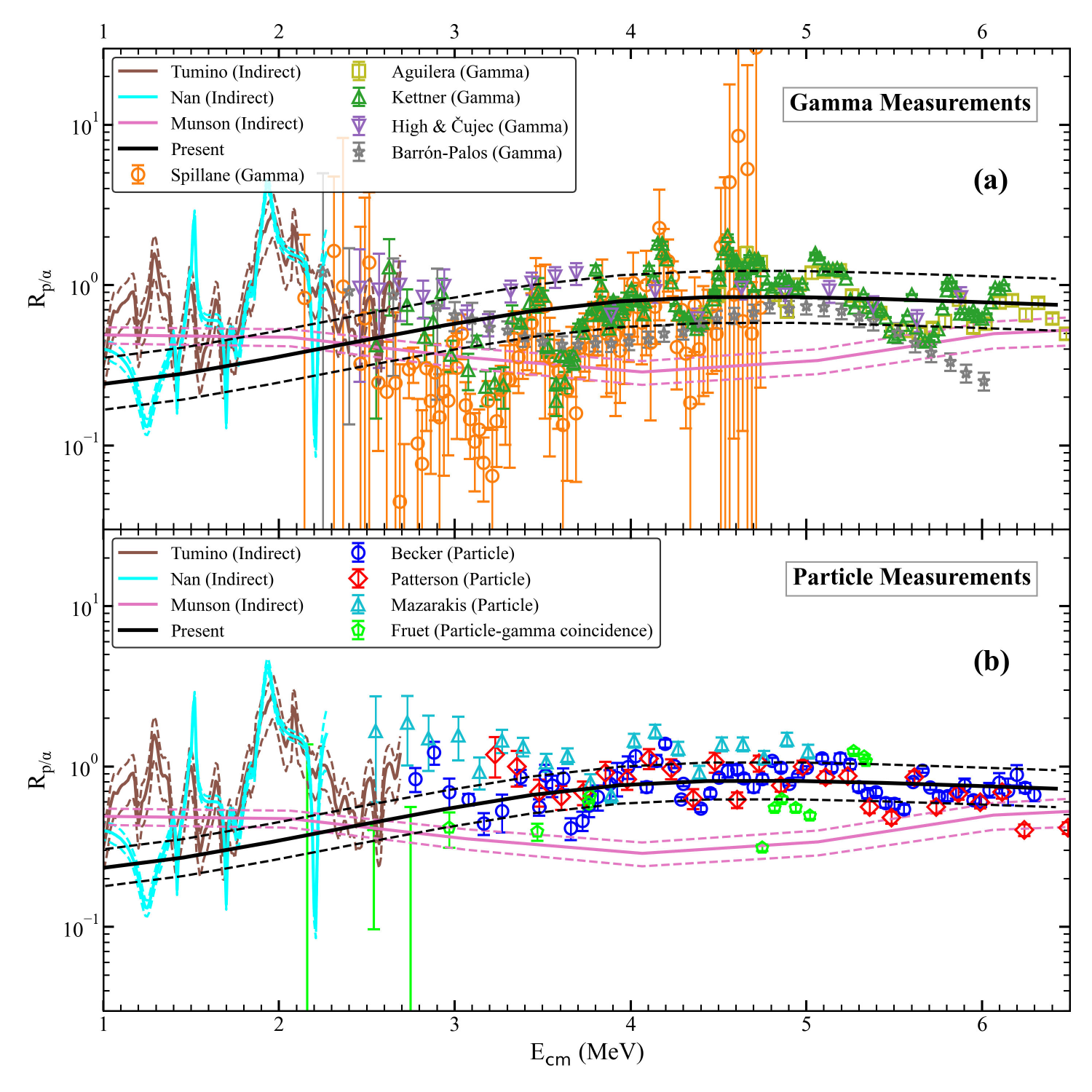}
	\caption{The proton to alpha channel ratios by direct measurements of Patterson $\emph{et al.}$ \cite{patterson1969}, Mazarakis and Stephens \cite{mazarakis1973}, High and {\v{C}}ujec \cite{high1977}, Kettner $\emph{et al.}$ \cite{kettner1977,kettner1980}, Becker $\emph{et al.}$ \cite{becker1981}, Aguilera $\emph{et al.}$ \cite{aguilera2006}, Barr{\'o}n-Palos $\emph{et al.}$ \cite{barron2006}, Spillane $\emph{et al.}$ \cite{spillane2007}, Fruet $\emph{et al.}$ \cite{fruet2020}, and by indirect methods of Munson $\emph{et al.}$ \cite{munson2017} , Nan \emph{et al.}\cite{nan2025} , and Tumino $\emph{et  al.}$ \cite{tumino2018}. All the measurement data have been corrected for their missing channels by applying the TALYS branching ratios of Li \emph{et al.}~\cite{li2020modified}. The black solid and dashed lines in panel (a) or (b) represent the recommended mean and +/-1$\delta$ errors based on our analysis from gamma or particle dataset (fit parameters are listed in Table~\ref{tab:exp_talys_fit}), respectively.}   
	\label{fig_p_a_ratio}
\end{figure*}

Figure~\ref{fig_p_a_ratio} presents a comprehensive comparison of the proton-to-alpha cross-section ratio for the
$^{12}$C+$^{12}$C fusion reaction obtained from different experimental approaches, including both direct and indirect measurements. All direct- and indirect-measurement data have been corrected for their missing reaction channels following the prescription of Li \emph{et al.}~\cite{li2020modified}. 
The pronounced resonance structures in the $^{12}$C+$^{12}$C fusion reaction produce strong fluctuations in $R_{p/\alpha}$, making it difficult to determine the ratio reliably at astrophysical energies.

Since the statistical model is expected to describe only the averaged behavior of the branching ratio, a purely theoretical prediction of the absolute proton-to-$\alpha$ cross-section ratio, $R_{p/\alpha}(E)$, remains challenging. 
This difficulty mainly arises from the presence of complex molecular resonances that are not fully incorporated in standard statistical-model calculations. To quantify the deviation between experimental data and statistical-model predictions, we define the ratio
\begin{equation}
R = \frac{(R_{p/\alpha}(E))_{\rm exp}}{f \times (R_{p/\alpha}(E))_{\rm Talys}},
\label{eq:R_exp_talys}
\end{equation}
where $R_{p/\alpha}(E)_{exp}$ and $R_{p/\alpha}(E)_{Talys}$ represent the experimental ratios obtained from experiments and the theoretical ratio predicted by TALYS , respectively, and $f$ represents the scaling factor applied to adjust the TALYS theoretical calculations.

To incorporate the statistical uncertainties of individual data points, we employ a Monte Carlo resampling technique. For each experimental point, 1000 random samples are generated according to its Gaussian probability distribution, $N(\mu_i, \sigma_i^2)$. This parametric bootstrap approach naturally weights the dataset: high-precision measurements produce tightly clustered samples, whereas less precise data generate broader distributions. The resulting ensemble is fitted with a Log-Normal distribution. It should be noted that the data are binned with equidistant intervals in the logarithmic space ($\ln R$). Consequently, the Jacobian factor $1/R$ is intrinsically absorbed, and the distribution in this logarithmic space follows a standard Gaussian profile, which is fitted using 
\begin{equation}
L(R) = A \exp\left[-\frac{(\ln R-\mu )^2}{2\delta^2}\right],
\label{eq:lognormal}
\end{equation}
where $A$ is a normalization constant, and $\mu$ and $\delta$ are the mean and standard deviation of $\ln R$, respectively. By varying the scaling factor $f$, the value of the ratio $R$ (defined in Eq. \eqref{eq:R_exp_talys}) can be adjusted. An appropriate choice of $f$ allows us to normalize the experimental ratio $(R_{p/\alpha}(E))_{\rm exp}$ to the scaled TALYS theoretical prediction $f \times (R_{p/\alpha}(E))_{\rm Talys}$, shifting the central value of $R$ to unity. Consequently, the mean of the logarithmic ratio becomes zero ($\ln 1 = 0$), corresponding to a central value $\mu \approx 0$ for the Log-Normal distribution of $\ln R$. For each dataset, $f$ was systematically optimized such that $\mu$ converged to zero within a tolerance of $\pm 0.002$. Furthermore, utilizing the standard deviation $\delta$ obtained from the Log-Normal fit of $\ln R$, we define the $1\delta$ confidence interval as $[ e^{\mu-\delta} \approx e^{-\delta}, e^{\mu+\delta}\approx e^{\delta} ]$. Therefore, the upper and lower limits of $f$ are estimated as $fe^{\delta}$ and $fe^{-\delta}$, respectively.

The scaled ratio, $f \times R_{p/\alpha}(E)_{Talys}$, is used to predict the ratio down to astrophysical energies, while the $\delta$ obtained from the fit is used to estimate the deviation between the true ratio and the TALYS prediction.

Different methods and different measurements bring different systematic uncertainties. In the following sections, we group the evaluations of $R_{p/\alpha}(E)_{exp}$  into three categories, measurements with charged particle spectroscopy, measurements with $\gamma$-ray spectroscopy, and indirect measurements. $f$ and $+/-\delta$ obtained for each category are compared with each other. Based on the comparison, the best value of $f$ is recommended together with the recommended deviation ($+/-\delta$).

\subsection{Fit to charged particle data}

As shown in Fig. 11 of Ref. \cite{li2016}, at higher energies (4.0 MeV<$E_{\rm cm}<$6.0 MeV), direct charged-particle measurements~\cite{patterson1969,mazarakis1973,becker1981} are generally consistent. However, below $\sim 4$ MeV, both the proton- and $\alpha$-channel $S^*$ values reported by Patterson and Mazarakis remain systematically higher than those of Becker. If these discrepancies originate from systematic uncertainties, such as those associated with target thickness, beam-current integration, and the absolute calibration of detection efficiencies, the experimental $R_{p/\alpha}$ ratio can effectively cancel most of these uncertainties.

We excluded the data of Mazarakis ~\cite{mazarakis1973} from our analysis due to a well-documented systematic energy offset of approximately 100 keV \cite{aguilera2006} and suspected beam-induced background from deuterium contaminants at low energies. 

To establish a robust experimental trend for the proton-to-alpha branching ratio $R_{p/\alpha}$, we performed a statistical comparison between the TALYS prediction and the direct particle data. As detailed in Table~\ref{tab:exp_talys_fit}, our primary analysis focuses on the energy window of $E_{\rm cm} = 3.5 - 6.4$ MeV. This interval is specifically chosen to avoid the deep sub-Coulomb regime ($E_{\rm cm} < 3.5$ MeV), where the degradation of the signal-to-noise ratio and significant background interference would otherwise introduce disproportionate weight and potential bias into the statistical inference.

The averaged ratio $R_{p/\alpha}$ was extracted using a Log-Normal distribution defined by Eq. \eqref{eq:lognormal}. Because the experimental uncertainties for cross-section ratios are predominantly multiplicative and the observables are strictly non-negative, the Log-Normal distribution provides a more rigorous statistical representation than a conventional Gaussian distribution.

\begin{figure}[tb]
    \centering
    \includegraphics[width=\columnwidth]{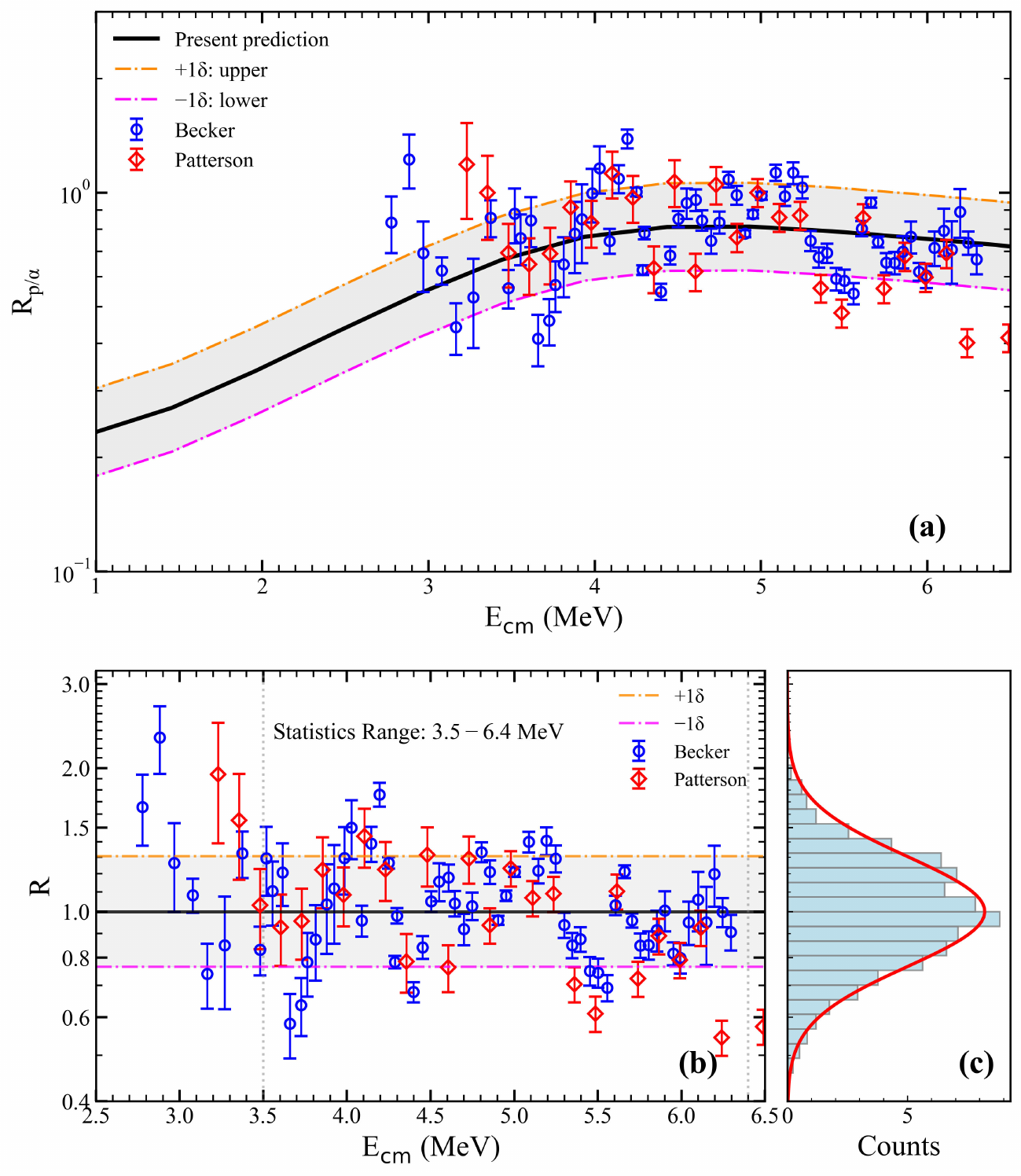}

 \caption{(a) Experimental ratio $R_{p/\alpha}$ compared with the present prediction and its $\pm1\delta$ uncertainty band. The central line corresponds to a scaling factor of 0.690, with uncertainty bounds of 0.767 and 1.303. Experimental data from Patterson \emph{et al.} and Becker \emph{et al.} are shown for comparison. (b) Ratio of experimental data to the scaled TALYS prediction, illustrating the extraction of the normalization factor. (c) Distribution of the ratio values shown in panel (b), together with a Log-Normal fit.}
\label{fig_particle_p_a_ratio_prediction}
\end{figure}

The results of this fitting procedure are comprehensively illustrated in Fig.~\ref{fig_particle_p_a_ratio_prediction}. The panel (b) of Fig.~\ref{fig_particle_p_a_ratio_prediction} shows the ratio of the experimental data to the scaled TALYS prediction, and panel (c) displays the corresponding Log-Normal distribution of these ratios. By fitting the datasets of Becker and Patterson within $E_{\rm cm} = 3.5 - 6.4$ MeV, we obtain an optimal scaling factor of $f = 0.690$. The statistical analysis yields a logarithmic standard deviation of $\delta = 0.265$. Consequently, the $1\delta$ uncertainty band is defined by multiplicative factors of $e^{-\delta} = 0.767$ (lower limit) and $e^{+\delta} = 1.303$ (upper limit). 

The scaled TALYS prediction, along with this precisely defined uncertainty band, is plotted against the experimental data in Fig.~\ref{fig_particle_p_a_ratio_prediction}(a), demonstrating excellent overall agreement within the selected energy window. Furthermore, Table~\ref{tab:exp_talys_fit} summarizes these fit parameters alongside those derived from other measurement techniques ($\gamma$-ray and indirect measurements), highlighting that the direct charged-particle data provide a tightly constrained averaged-trend for calibrating the theoretical $R_{p/\alpha}$ ratio.

\begin{table*}[htbp]
	\centering
	\setlength{\tabcolsep}{3pt} 
	\renewcommand{\arraystretch}{1.3}
	\caption{Log-Normal fit results of experimental $R_{p/\alpha}$ ratios normalized to present prediction.}
	\label{tab:exp_talys_fit}
	\begin{threeparttable}  
	\begin{tabular}{c p{8.6cm}ccccc} 
		\toprule[0.75pt]
		\bfseries No. & \bfseries Fit dataset & \bfseries Fit range & \bfseries $f$ & \bfseries $\delta$ & \bfseries $e^{-\delta}$(lower) & \bfseries $e^{+\delta}$(upper) \\
		\midrule[0.5pt]
		1 & \textbf{Particle}: Becker \cite{becker1981}, Patterson \cite{patterson1969} & 3.5-6.4 MeV  & 0.690 & 0.265 & 0.767 & 1.303 \\
		\hline
		2 & \textbf{Gamma}: Aguilera \cite{aguilera2006}, Kettner \cite{kettner1977,kettner1980}, High and {\v{C}}ujec \cite{high1977} & 3.7-6.4 MeV & 0.715 & 0.373 & 0.689 & 1.452 \\
		\hline
		3 & \multirow{2}{=}{\textbf{Particle \& Gamma}: Becker \cite{becker1981}, Patterson \cite{patterson1969}, Aguilera \cite{aguilera2006}, Kettner \cite{kettner1977,kettner1980}, High and {\v{C}}ujec \cite{high1977}} 
		& 3.5-6.4 MeV & 0.698\tnote{\S} & 0.345 & 0.708 & 1.412 \\
		\cline{3-7}
		& & 3.7-6.4 MeV & 0.698 & 0.330 & 0.719 & 1.391 \\
		\hline
		4 & \textbf{Indirect}: Nan \cite{nan2025}  & 1.0-3.0 MeV  & 1.319 & 0.773 & 0.462 & 2.166 \\
		\hline
		5 & \textbf{Indirect}: Nan \cite{nan2025}, Tumino \cite{tumino2018}  & 1.0-3.0 MeV  & 1.443 & 0.698 & 0.498 & 2.010 \\
		\bottomrule[0.75pt]
	\end{tabular}
    \begin{tablenotes}    
        \footnotesize               
        \item[\S] Recommended value 0.698 for scaling factor $f$.         
      \end{tablenotes}   
    \end{threeparttable}     
\end{table*}

It is worth noting that extracting the total partial cross sections from experimental data requires correcting for unobserved weak decay channels (e.g., highly excited states), for which we utilized the statistical model estimations from Li \emph{et al.} \cite{li2020modified}. While this introduces a degree of model dependence into the ``experimental" ratio, these unobserved channels generally account for a minor fraction of the total reaction yield. Consequently, the extracted central scaling factor $f$ remains predominantly constrained by the direct measurements of the dominant decay branches.

\subsection{Fit to $\gamma$-ray spectroscopy data and global analysis}

In addition to charged-particle measurements, $\gamma$-ray spectroscopy provides an alternative method for extracting partial cross sections. As shown in Fig.~\ref{fig_p_a_ratio}, the ratios derived from $\gamma$-ray data exhibit much stronger fluctuations, with $R_{p/\alpha}$ varying drastically between approximately 0.2 and 2.0, whereas the ratios extracted from charged-particle measurements fluctuate within a narrower range of about 0.5 to 1.5. Since $\gamma$-ray spectroscopy misses more reaction channels than charged-particle measurements, the extracted ratios are more sensitive to the complex cluster structures and molecular resonances inherent in the $^{12}\mathrm{C}+^{12}\mathrm{C}$ system near and below the Coulomb barrier, leading to significantly larger fluctuations.

\begin{figure}[tb]
    \centering
    \includegraphics[width=\columnwidth]{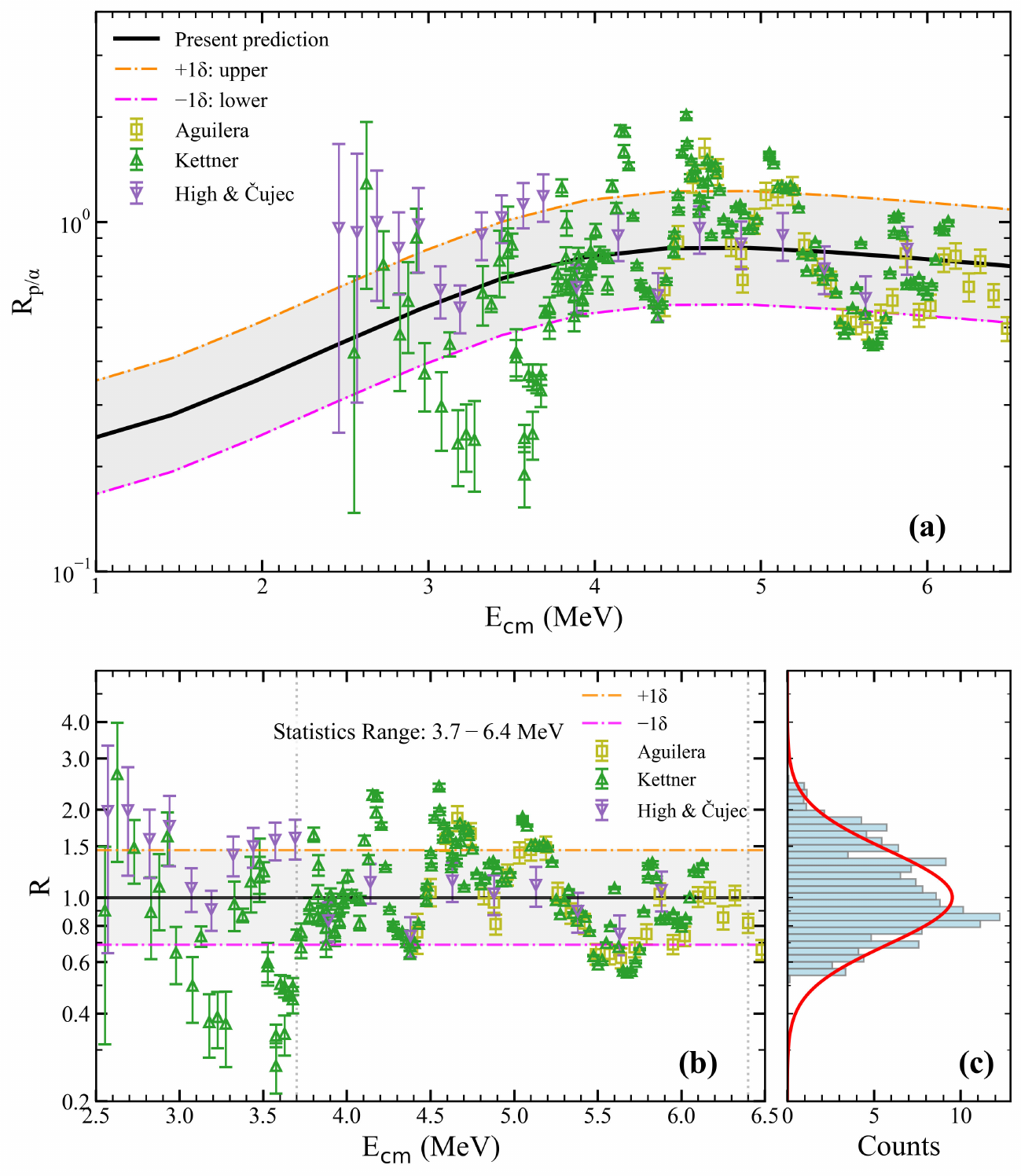}
    \caption{(a) Experimental ratio $R_{p/\alpha}$ compared with the present prediction and its $\pm1\delta$ uncertainty band. The data are taken from Aguilera, Kettner, and High and {\v{C}}ujec. (b) Ratio between the experimental data and the TALYS calculation scaled by a factor of 0.715 in the energy range $E_{\mathrm{cm}}=3.7$--$6.4$ MeV. (c) Log-Normal distribution fit of the ratio values shown in panel (b).}
    \label{fig:ratio_combined_gamma}
\end{figure}

We applied our Log-Normal distribution defined in Eq. \eqref{eq:lognormal} to fit the TALYS prediction to the datasets of Aguilera \emph{et al.}~\cite{aguilera2006}, Kettner \emph{et al.}~\cite{kettner1977,kettner1980}, and High and {\v{C}}ujec~\cite{high1977}. As summarized in Table~\ref{tab:exp_talys_fit} and depicted in Fig.~\ref{fig:ratio_combined_gamma}, fitting these datasets within the reliable energy window of $E_{\mathrm{cm}} = 3.7 - 6.4$ MeV yields an optimal scaling factor of $f = 0.715$. The statistical analysis of the data-to-theory ratios gives a logarithmic standard deviation corresponding to multiplicative uncertainty bounds of $e^{-\delta} = 0.689$ (lower limit) and $e^{+\delta} = 1.452$ (upper limit). Remarkably, this central scaling factor ($f = 0.715$) is in excellent agreement with the factor derived independently from the charged-particle data ($f$ = 0.690). The noticeably wider uncertainty band, however, correctly encapsulates the large intrinsic resonance fluctuations present in the $\gamma$-ray channels.

\begin{figure}[tb]
    \centering
    \includegraphics[width=\columnwidth]{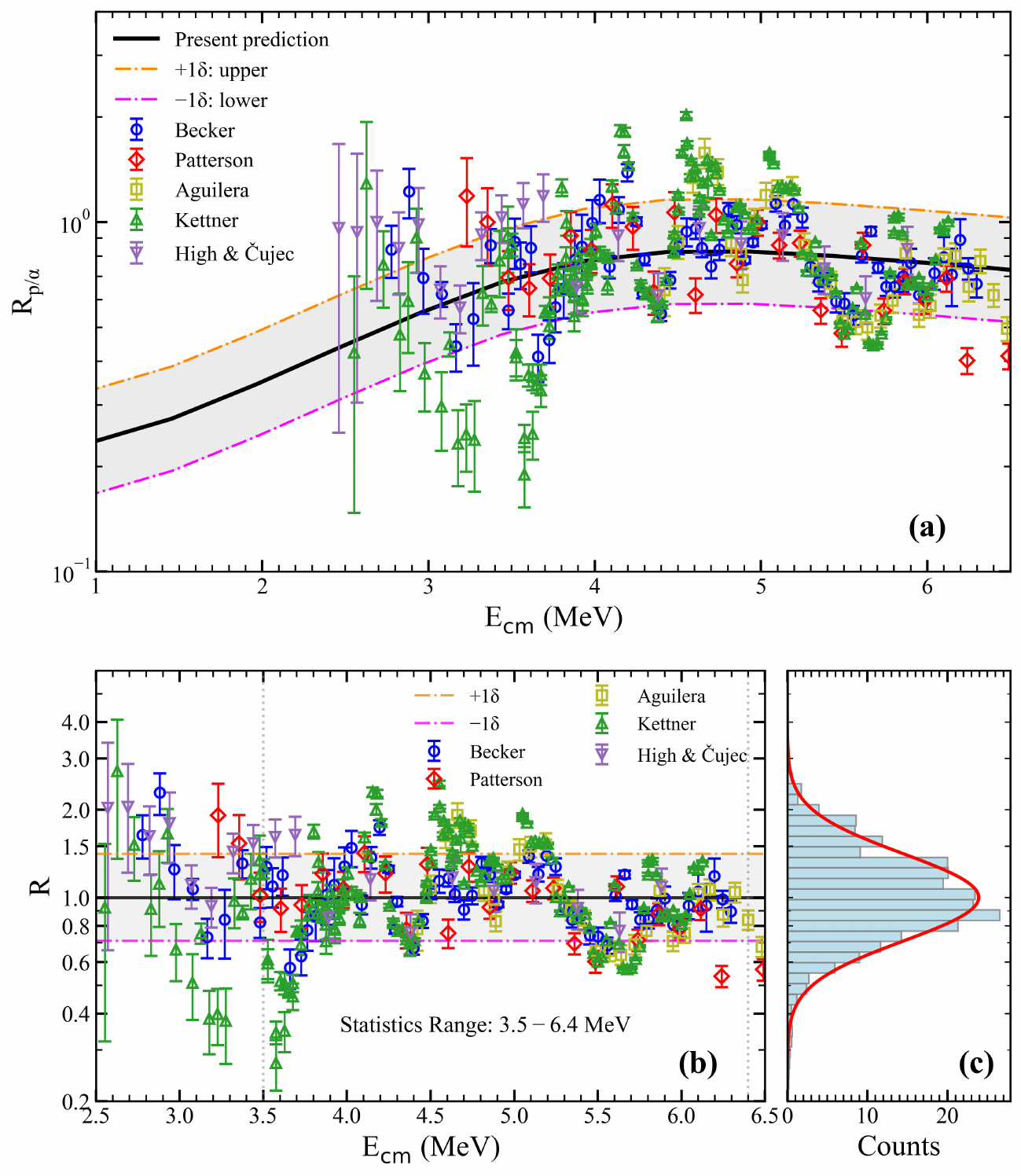}
    \caption{(a) Experimental ratio $R_{p/\alpha}$ compared with the present prediction and its $\pm1\delta$ uncertainty band. The comprehensive dataset includes measurements from Patterson, Becker, Aguilera, Kettner, and High and {\v{C}}ujec. (b) Ratio between the experimental data and the TALYS calculation scaled by a global factor of 0.698 in the energy range $E_{\mathrm{cm}}=3.5$--$6.4$ MeV. (c) Log-Normal distribution fit of the ratio values shown in panel (b).}
    \label{fig:ratio_combined_particle_gamma}
\end{figure}

Given the excellent systematic agreement in the central scaling factors derived independently from the primary charged-particle and $\gamma$-ray datasets, we construct a comprehensive global fit to establish the final theoretical averaged-trend. By combining the data from Patterson, Becker, Aguilera, Kettner, and High and {\v{C}}ujec in the energy range $E_{\mathrm{cm}} = 3.5 - 6.4$ MeV, we obtain a global optimal scaling factor of $f$ = 0.698, which is the recommended value from present work, with an uncertainty band defined by $e^{-\delta} = 0.708$ and $e^{+\delta} = 1.412$ (see Table~\ref{tab:exp_talys_fit} and Fig.~\ref{fig:ratio_combined_particle_gamma}). This combined fit solidifies the normalization of the TALYS calculation, proving that the ratio $R_{p/\alpha}$ converges consistently regardless of the distinct systematic errors inherent to particle or $\gamma$-ray detection techniques. At sub-barrier energies, a value of $f \approx 0.7$ suggests that the standard OMP may underestimate the $\alpha$-particle penetrability for the $^{12}$C+$^{12}$C system.

\subsection{Comparison with indirect measurements}

Deep inside the Coulomb barrier ($E_{\rm cm} < 2.5$ MeV), reliable direct measurements become extremely scarce, leaving only indirect data~\cite{munson2017,tumino2018,nan2025}. 
The Trojan Horse Method (THM) employed by Tumino \emph{et al.}~\cite{tumino2018} was originally analyzed within the plane-wave approximation. Although subsequent corrections using the distorted-wave approximation led to a substantially different energy dependence of the extracted cross sections, the resulting $R_{p/\alpha}$ ratio remained unaffected.
Munson \emph{et al.}~\cite{munson2017} populated the ${}^{24}$Mg compound nucleus through inelastic scattering of 40 MeV $\alpha$ particles. The resulting $R_{p/\alpha}$ ratio differs significantly from that obtained in direct measurements. This discrepancy arises because the surrogate reaction mechanism populates a spin-parity distribution that is substantially different from that produced in sub-barrier ${}^{12}$C+${}^{12}$C fusion. Consequently, the extracted decay branching ratios are not directly applicable to quiescent stellar carbon burning. Future indirect measurements must therefore constrain the populated spin-parity distribution in order to reliably determine the proton-to-alpha competition deep within the Gamow window. 
In the study of $p({}^{23}{\rm Na},\alpha){}^{20}{\rm Ne}$ by Nan \emph{et al} \cite{nan2025}, excitation functions for the proton and $\alpha$ channels were measured using a thick-target inverse kinematics approach combined with $\gamma$-particle coincidence techniques. A multi-channel $R$-matrix analysis revealed a series of discrete 0$^+$ and 2$^+$ resonances in the relevant energy region and provided branching ratios for the dominant decay channels across the Gamow window. These results highlight the strong energy dependence and fluctuations of the channel competition, which have important implications for observables such as the $R_{p/\alpha}$ ratio and the resulting astrophysical reaction rates.

\begin{figure}[tb]
    \centering
    \includegraphics[width=\columnwidth]{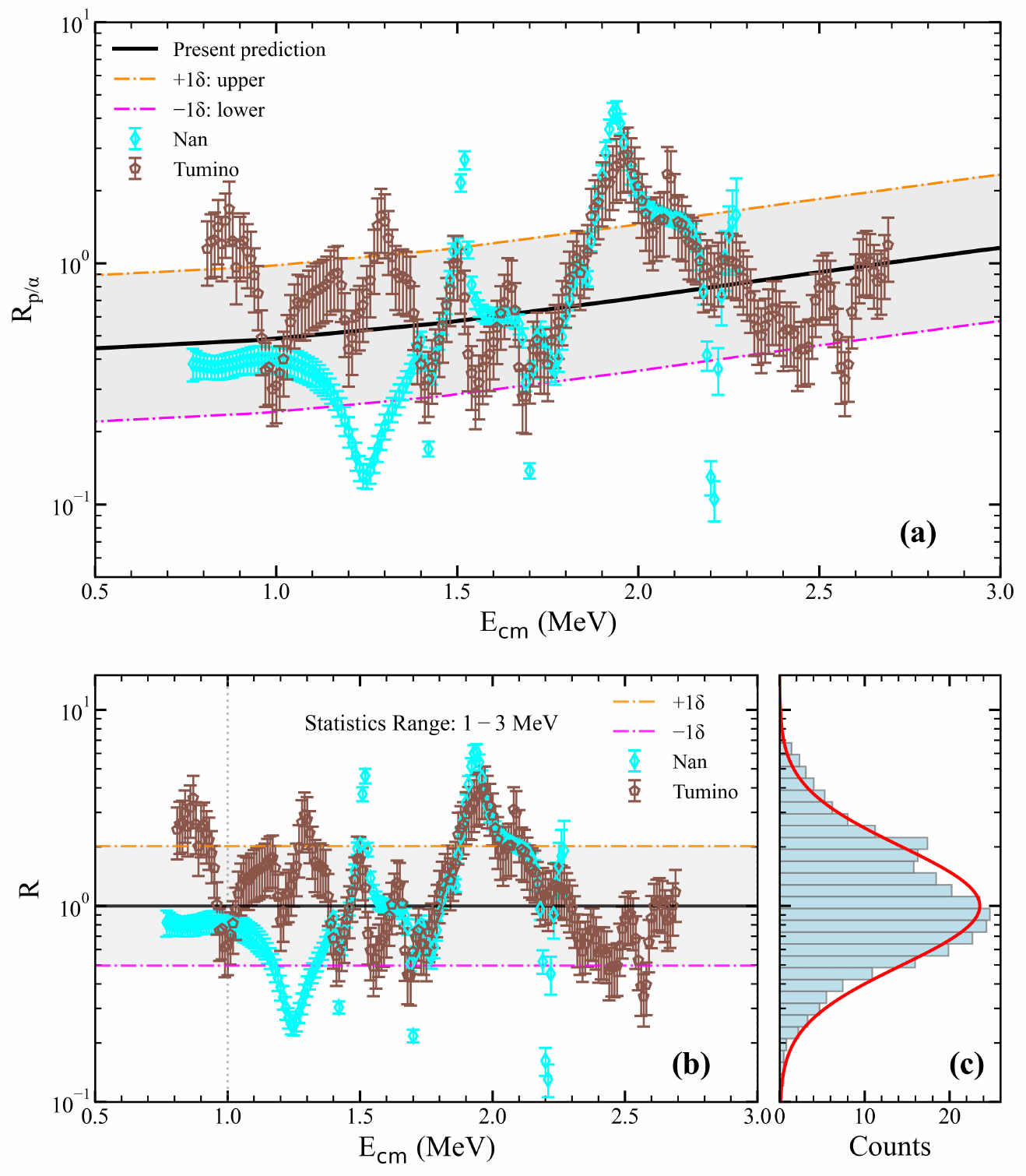}
    \caption{
    (a) Experimental ratio $R_{p/\alpha}$ compared with the present prediction and its $\pm1\delta$ uncertainty band.
    The data are taken from Ref.\cite{tumino2018,nan2025}.
    (b) Ratio between the experimental data and the TALYS calculation scaled by a factor of 1.443 in the energy range
    $E_{\mathrm{cm}}=1$--$3$ MeV.
    (c) Distribution of the ratio values shown in panel (b), together with a Log-Normal fit.
    }   
    \label{fig:ratio_combined_indirect}
\end{figure}

Figure~\ref{fig:ratio_combined_indirect} presents the combined analysis of the THM datasets reported by Tumino \emph{et al.}~\cite{tumino2018} and Nan \emph{et al.}~\cite{nan2025}. As the experiments only resolved partial transitions to specific low-lying states (predominantly the $p_0$, $p_1$, $\alpha_0$, and $\alpha_1$ channels). Consequently, the data points were corrected using the statistical model estimations of the missing-channel branching ratios provided by Li \emph{et al.}~\cite{li2020modified}. 

By applying our Log-Normal distribution to fit these corrected indirect datasets against the TALYS prediction within the energy range of $E_{\mathrm{cm}} = 1 - 3$ MeV, we obtain an scaling factor of $f = 1.443$. As depicted in Fig.~\ref{fig:ratio_combined_indirect}(c), the resulting Log-Normal distribution exhibits a significantly broad dispersion. Most importantly, this indirectly derived scaling factor ($f = 1.443$) deviates substantially—by roughly a factor of two—from the  averaged-trend established by the direct charged-particle and gamma-ray measurements ($f = 0.698$). This striking discrepancy highlights the persistent systematic differences between direct and indirect techniques and calls for further theoretical investigations to clarify their origin.

\subsection{Prediction of the $R_{p/\alpha}$ ratio within the Gamow window}

While our predicted averaged $R_{p/\alpha}$ ratio is generally consistent with the CF88 \cite{caughlan1988} value ($\sim 0.8$) above $E_{\rm cm}=4.0$ MeV, it decreases significantly as the energy extends deeper into the Gamow window. The strong suppression of the proton channel relative to the $\alpha$ channel at low energies is primarily governed by the availability of the decay channels.

\begin{figure*}[!htbp]
	\centering
	\includegraphics[width=0.92\textwidth]{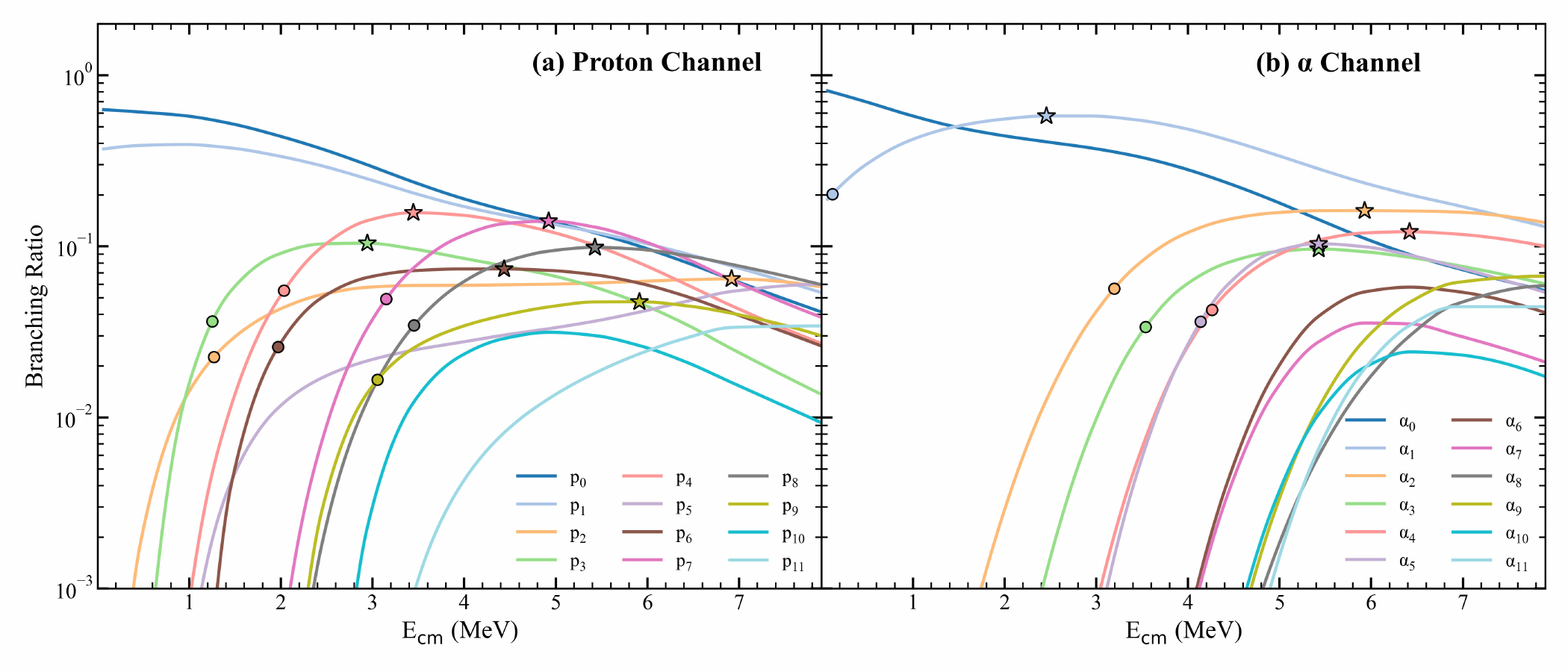}
	\caption{The branching ratios for the different channels of $^{12}$C($^{12}$C,p)$^{23}$Na and $^{12}$C($^{12}$C,$\alpha$)$^{20}$Ne reactions calculated by Li \emph{et al.}~\cite{li2020modified}.  The cut off energies at which the branching ratio of specific single channel($p_2$, $p_3$, $p_4$, $p_6$, $p_7$, $p_8$, $p_9$, $\alpha_1$, $\alpha_2$, $\alpha_3$, $\alpha_4$, $\alpha_5$) drops from the maximum value (B$_{max}$, marked with stars) down to 35\% of B$_{max}$ with decreasing energy are marked with circles.}
	\label{fig_talys_branching_ratio}
\end{figure*}

\begin{figure}[!htbp]
	\centering
	\includegraphics[width=0.5\textwidth]{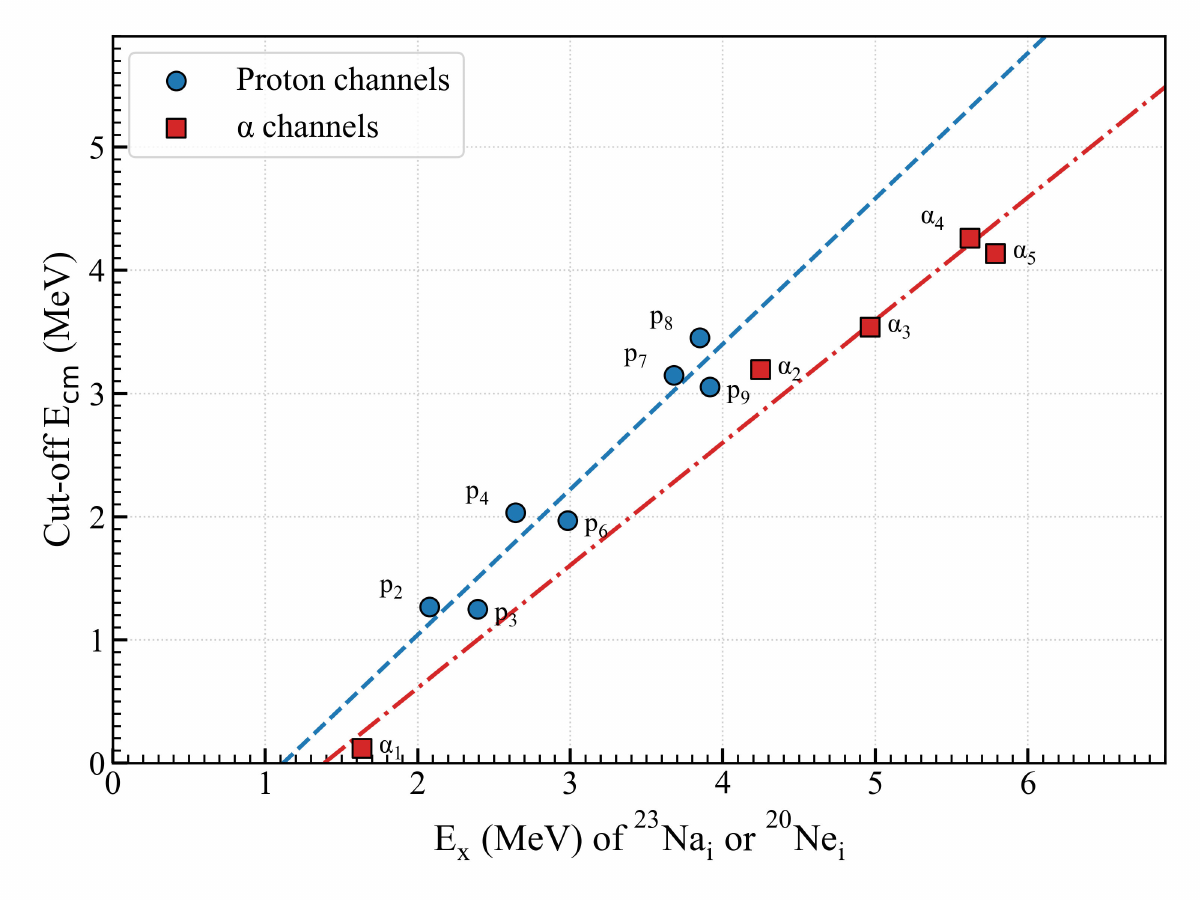}
	\caption{The cut off energies of specific single channel($p_2$, $p_3$, $p_4$, $p_6$, $p_7$, $p_8$, $p_9$, $\alpha_1$, $\alpha_2$, $\alpha_3$, $\alpha_4$, $\alpha_5$)  versus the excited energy ($E_{\rm x}$) of related $^{23}$Na or $^{20}$Ne. The solid round dots are the values of proton channels including $p_2$, $p_3$, $p_4$, $p_6$, $p_7$, $p_8$, $p_9$; the solid square dots are the values of $\alpha$ channels including $\alpha_1$, $\alpha_2$, $\alpha_3$, $\alpha_4$, $\alpha_5$. These values come from the previous calculation by Li $\emph{et al.}$ \cite{li2020modified} (Fig.~\ref{fig_talys_branching_ratio}). The dashed and dashdot lines are the linear fitting for displayed points of proton and $\alpha$ channels, respectively. }
	\label{fig_cut_off_ecm}
\end{figure}

As the center-of-mass energy decreases, the number of energetically accessible proton-emission channels diminishes much more rapidly than that of the $\alpha$-emission channels. To illustrate this behavior systematically, we investigated the critical cut-off energies at which the branching ratios for specific highly excited states (e.g., $p_2$ to $p_9$ and $\alpha_2$ to $\alpha_5$) decrease to 35\% of their respective maximum values (Fig.~\ref{fig_talys_branching_ratio}). As shown in Fig.~\ref{fig_cut_off_ecm}, at $E_{\rm cm} \sim 4.0$ MeV, where $R_{p/\alpha} \sim 0.78$, the energetically open proton-emission channels extend from $p_0$ to $p_9$, whereas the open $\alpha$-emission channels are limited mainly to $\alpha_0$, $\alpha_1$ and $\alpha_2$.

As the energy decreases further from 4.0 MeV down to 0.5 MeV, the proton-emission channels close sequentially, whereas the $\alpha_0$ and $\alpha_1$ channels remain open down to $E_{\rm cm}=0$ MeV. The $Q$ value for the $^{12}$C($^{12}$C,$\alpha$)$^{20}$Ne reaction is 4.62 MeV, substantially larger than the 2.24 MeV associated with the $^{12}$C($^{12}$C,$p$)$^{23}$Na reaction. This sequential closure of proton decay channels, originating from the intrinsically smaller $Q$ value of the proton channel, strongly suppresses the competitiveness of proton evaporation at stellar energies. Such a mechanism naturally leads to the pronounced decline of the $R_{p/\alpha}$ ratio predicted within the Gamow window.

\section{New Reaction-Rate Ratio, $\langle \sigma v \rangle_p / \langle \sigma v \rangle_\alpha$, and Its Impact on Astrophysical Models}

\begin{figure}[!htbp]
	\centering
	\includegraphics[width=0.49\textwidth]{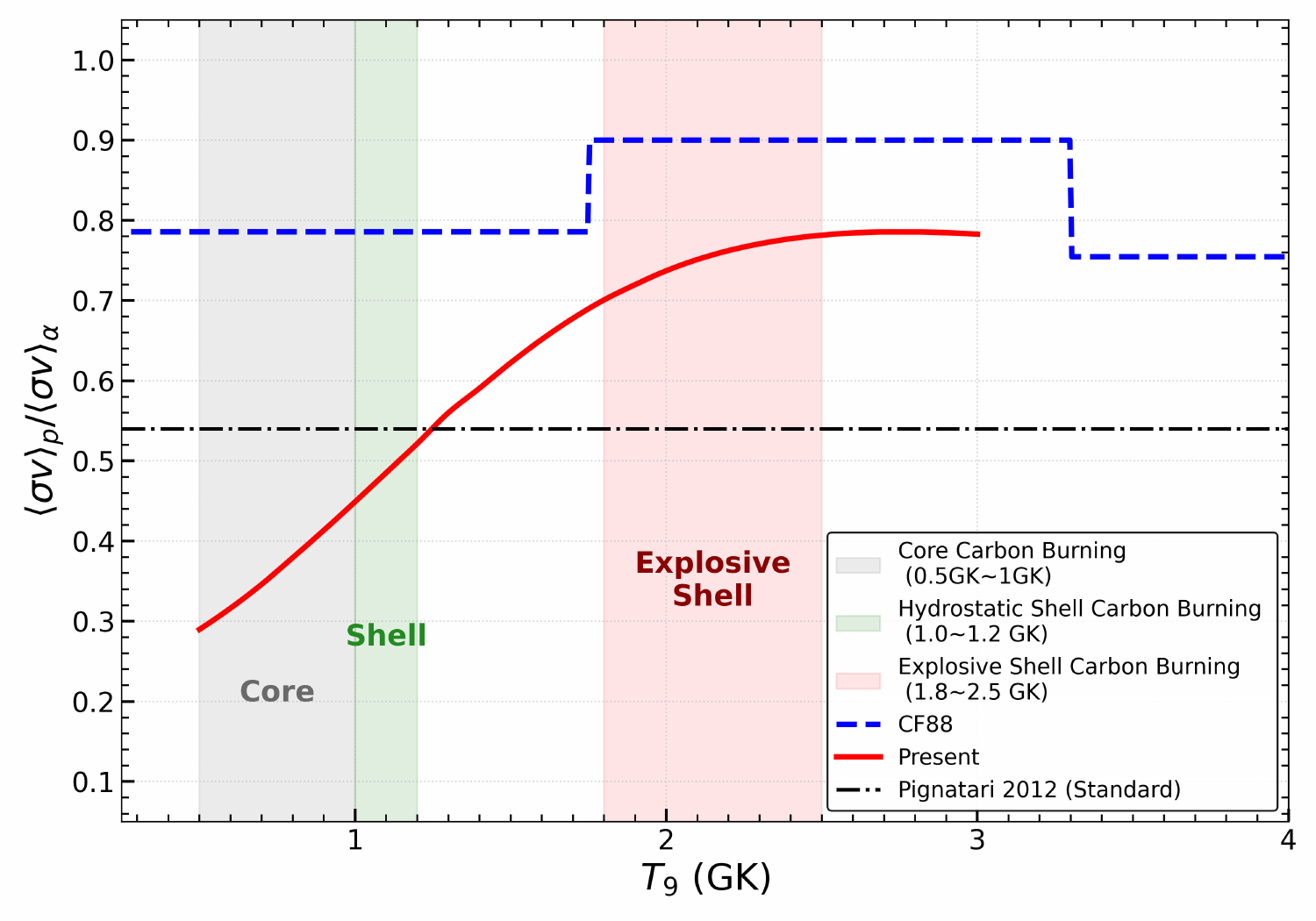}
	\caption{Temperature dependence of the reaction rate ratio $\langle \sigma v \rangle_p / \langle \sigma v \rangle_\alpha$ from present work (solid curve), compared with the recommended values from CF88 \cite{caughlan1988} (dashed lines) and Pignatari $\emph{et al.}$ \cite{pignatari2012} (0.35/0.65 $\approx$ 0.54, dash-dotted line). The different stages of stellar carbon burning are indicated by colored shaded regions: core carbon burning ($0.5 \sim 1.0$ GK, gray), hydrostatic shell carbon burning ($1.0 \sim 1.2$ GK, green), and explosive shell carbon burning ($1.8 \sim 2.5$ GK, red).}
	\label{fig_p_a_ratio_rate}
\end{figure}

We re-calculated the reaction rate of $^{12}{\rm C}(^{12}{\rm C},p){}^{23}{\rm Na}$ and $^{12}{\rm C}(^{12}{\rm C},\alpha){}^{20}{\rm Ne}$ by assuming a constant $S^*$=3$\times10^{16}$ MeV$\cdot$barn recommended by CF88 for the $^{12}$C+$^{12}$C fusion reaction and the $R_{p/\alpha}$ ratio established in this work. The reaction rate ratio, $\langle \sigma v \rangle_p / \langle \sigma v \rangle_\alpha$, is shown in Fig.~\ref{fig_p_a_ratio_rate}. The present work reveals a pronounced temperature dependence of this ratio, which deviates significantly from the widely used CF88 model \cite{caughlan1988}. 

While the CF88 compilation assumes a nearly constant $\langle \sigma v \rangle_p / \langle \sigma v \rangle_\alpha$ ratio of $\sim 0.79$ over most astrophysical temperatures, our results exhibit a strong temperature dependence. The ratio increases from 0.29 at $T_9=0.5$ to 0.45 at $T_9=1.0$, 0.52 at $T_9=1.2$, and 0.70 at $T_9=1.8$, reaching $\sim$0.78 at $T_9=2.5$, and then approaches a nearly constant value at higher temperatures. Compared with CF88, our lower ratios at $T_9 \lesssim 1.2$ imply a stronger dominance of the $\alpha$ channel during hydrostatic carbon burning, potentially modifying the seed distribution for subsequent weak $s$-process nucleosynthesis. At explosive carbon-burning temperatures ($T_9 \gtrsim 2$), the ratio gradually approaches the CF88 prediction, indicating an increasing contribution from the proton-exit channel in high-temperature stellar environments.

Pignatari \emph{et al.}~\cite{pignatari2012} investigated the impact of uncertainties in the carbon-burning reaction channels on stellar nucleosynthesis. Three different values of the reaction-rate ratio, $\langle \sigma v \rangle_p / \langle \sigma v \rangle_\alpha = 0.54$, 19, and 0.05, were adopted to explore the influence of the proton-to-$\alpha$ competition on the abundances of nuclei that serve as the primary fuel for subsequent stellar evolutionary stages. 
Their calculations demonstrated that the competition between the proton and $\alpha$ channels in the $^{12}$C+$^{12}$C reaction plays a crucial role in determining the nucleosynthesis flow during carbon burning. In the proton-dominated scenario, proton-induced reaction chains and neutron production through the ${}^{13}$C($\alpha,n$)${}^{16}$O reaction are significantly enhanced, leading to stronger weak $s$-process nucleosynthesis and increased abundances of odd-$Z$ nuclei such as ${}^{23}$Na and ${}^{26}$Al. In contrast, the strongly $\alpha$-dominated case suppresses proton-induced reaction flows and shifts the nucleosynthesis path toward $\alpha$-capture products. Our predicted averaged value of $\langle \sigma v \rangle_p / \langle \sigma v \rangle_\alpha$ lies much closer to the standard value of 0.54 than to either of the two extreme scenarios, as shown in Fig. \ref{fig_p_a_ratio_rate}. We suggest that incorporating these temperature-dependent ratios into stellar and galactic chemical-evolution models will lead to more reliable predictions of the $^{23}\mathrm{Na}/^{20}\mathrm{Ne}$ abundance ratio in massive stars.

As recently demonstrated by the calculations of De Gerónimo \emph{et al.}~\cite{DeGeronimo2024}, which were based on our recommended branching ratio, uncertainties in the $\langle \sigma v \rangle_p / \langle \sigma v \rangle_\alpha$ ratio have important consequences for the structural and chemical evolution of super-asymptotic giant branch (SAGB) stars and their descendants, ultramassive white dwarfs (UMWDs).
Since the proton and $\alpha$ channels predominantly produce $^{23}\mathrm{Na}$ and $^{20}\mathrm{Ne}$, respectively, the branching ratio directly determines the chemical composition of the stellar core after carbon burning. De Gerónimo \emph{et al.}~\cite{DeGeronimo2024} showed that the uncertainty in the branching ratio is one of the dominant sources of uncertainty in the final chemical structure of SAGB progenitors. In particular, a stronger $\alpha$ channel (corresponding to a lower $\langle \sigma v \rangle_p / \langle \sigma v \rangle_\alpha$ ratio) substantially enhances the central $^{20}\mathrm{Ne}$ abundance, with variations reaching up to $\sim17\%$. Such changes in the $^{20}\mathrm{Ne}/^{16}\mathrm{O}$ ratio affect the crystallization, cooling behavior, and pulsational properties of the resulting UMWDs. 

The branching ratio also influences the threshold mass for carbon ignition. Because the $\alpha$ channel releases more energy than the proton channel, a more dominant $\alpha$ branch increases the efficiency of carbon burning and allows carbon ignition to occur at slightly lower stellar masses. De Gerónimo \emph{et al.}~\cite{DeGeronimo2024} found that the minimum zero-age main-sequence (ZAMS) mass required to experience a carbon flash may decrease by approximately $1.5\%$ when the $\alpha$ channel dominates. In addition, the mass range in which carbon burning remains partially incomplete is also sensitive to the adopted $\langle \sigma v \rangle_p / \langle \sigma v \rangle_\alpha$ ratio.

We acknowledge that extrapolating the Hauser-Feshbach statistical model into the deep sub-Coulomb Gamow window ($E_{\rm cm} < 3$ MeV) encounters inherent physical limitations. In this regime, the level density of the compound nucleus $^{24}$Mg is relatively low, and the reaction mechanism is strongly modulated by discrete, structured molecular resonances rather than overlapping statistical states. Our physically calibrated TALYS prediction serves as an averaged statistical trend. In stellar environments, the reaction rate is an integral over the Gamow window. While individual discrete resonances will undoubtedly cause local fluctuations in the $R_{p/\alpha}$ ratio, their energy-averaged collective behavior is expected to follow this statistical trend. Most importantly, providing an energy-dependent, theoretically and experimentally constrained averaged trend of $\langle \sigma v \rangle_p / \langle \sigma v \rangle_\alpha$ represents a significant improvement over the arbitrary constant branching ratio historically adopted in standard astrophysical networks (e.g., CF88), providing a more realistic ``prior" for future R-matrix evaluations of specific discrete resonances.

%%*****************************************************************************************
\section{Summary}

In summary, we have investigated the energy dependence of the $^{12}\text{C}({}^{12}\text{C},p)^{23}\text{Na}$ to $^{12}\text{C}({}^{12}\text{C},\alpha)^{20}\text{Ne}$ branching ratio and its impact on carbon-burning nucleosynthesis. A careful re-examination of sub-barrier experimental data, reveals significant discrepancies at $E_{\text{cm}} < 3.0$ MeV. By focusing on the more consistent high-energy regime, we derived a new temperature-dependent branching ratio for astrophysical environments. Our results demonstrate that $R_{p/\alpha}$ is not a constant but evolves dynamically with temperature, and $\langle \sigma v \rangle_p / \langle \sigma v \rangle_\alpha$ decreases from nearly 0.78 at $T_9 = 2.5$ to 0.29 at $T_9 = 0.5$.

The deviation from the traditional CF88 constant ratio may have implications for the various stages of hydrostatic carbon burning and the subsequent formation of white dwarfs. While the present statistical framework cannot fully resolve the structures of isolated resonances, the temperature-dependent $\langle \sigma v \rangle_p / \langle \sigma v \rangle_\alpha$ averaged trend established in this work provides a physically motivated alternative to the widely adopted CF88 constant values. This energy-dependent trend serves as an updated reference for modeling stellar chemical evolution. Ultimately, determining the exact branching ratios still awaits highly precise direct measurements deep within the Gamow window.

%%*****************************************************************************************
\begin{acknowledgments}
This work was supported by the GuangDong Basic and Applied Basic Research Foundation (2026A1515011317), National Natural Science Foundation of China (11875329, 12335009, U1632142), and National Key Research and Development program (MOST 2022YFA1602304).
\end{acknowledgments}
%%\bibliography{branching_ratio}

\begin{thebibliography}{99}
	
	\bibitem{Gasques2005}
	L. R. Gasques, A. V. Afanasjev, E. F. Aguilera, et al., Phys. Rev. C, 72 (2): 025806 (2005)
	
	\bibitem{DeGeronimo2024}
	F. C. De Ger\'{o}nimo, M. M. Miller Bertolami, T. Battich, et al., Astrophys. J., 975 (2): 259 (2024)
	
	\bibitem{chieffi2025status}
	A. Chieffi, S. Courtin, R. J. Deboer, et al., Eur. Phys. J. A, 61 (12): 280 (2025)
	
	\bibitem{Wiescher2025Quantum}
	M. Wiescher, C. A. Bertulani, C. R. Brune, et al., Rev. Mod. Phys., 97 (2): 025003 (2025)
	
	\bibitem{patterson1969}
	J. R. Patterson, H. Winkler, C. S. Zaidins, Astrophys. J., 157: 367 (1969)
	
	\bibitem{pignatari2012}
	M. Pignatari, R. Hirschi, M. Wiescher, et al., Astrophys. J., 762 (1): 31 (2012)
	
	\bibitem{li2020modified}
	Y. J. Li, X. Fang, B. Bucher, et al., Chin. Phys. C, 44 (11): 115001 (2020)
	
	\bibitem{koning2005}
	A. J. Koning, S. Hilaire, M. C. Duijvestijn, in AIP Conf. Proc., Vol. 769 (AIP, 2005), p. 1154
	
	\bibitem{koning2023talys}
	A. Koning, S. Hilaire, S. Goriely, Eur. Phys. J. A, 59 (6): 1-85 (2023)
	
	\bibitem{hauser1952}
	W. Hauser, H. Feshbach, Phys. Rev., 87 (2): 366 (1952)
	
	\bibitem{hagino1999}
	K. Hagino, N. Rowley, A. T. Kruppa, Comput. Phys. Commun., 123: 143-152 (1999)
	
	\bibitem{mazarakis1973}
	M. G. Mazarakis, W. E. Stephens, Phys. Rev. C, 7 (4): 1280 (1973)
	
	\bibitem{high1977}
	M. D. High, B. \v{C}ujec, Nucl. Phys. A, 282 (1): 181-188 (1977)
	
	\bibitem{kettner1977}
	K. U. Kettner, H. Lorenz-Wirzba, C. Rolfs, et al., Phys. Rev. Lett., 38 (7): 337 (1977)
	
	\bibitem{kettner1980}
	K. U. Kettner, H. Lorenz-Wirzba, C. Rolfs, Z. Phys. A, 298 (1): 65-75 (1980)
	
	\bibitem{becker1981}
	H.-W. Becker, K. U. Kettner, C. Rolfs, et al., Z. Phys. A, 303 (4): 305-312 (1981)
	
	\bibitem{aguilera2006}
	E. F. Aguilera, P. Rosales, E. Martinez-Quiroz, et al., Phys. Rev. C, 73 (6): 064601 (2006)
	
	\bibitem{barron2006}
	L. Barr\'{o}n-Palos, E. F. Aguilera, J. Aspiazu, et al., Nucl. Phys. A, 779: 318-332 (2006)
	
	\bibitem{spillane2007}
	T. Spillane, F. Raiola, C. Rolfs, et al., Phys. Rev. Lett., 98: 122501 (2007)
	
	\bibitem{fruet2020}
	G. Fruet, S. Courtin, M. Heine, et al., Phys. Rev. Lett., 124 (19): 192701 (2020)
	
	\bibitem{munson2017}
	J. M. Munson, E. B. Norman, J. T. Burke, et al., Phys. Rev. C, 95 (1): 015805 (2017)
	
	\bibitem{nan2025}
	W. Nan, Y. Wang, J. Su, et al., Phys. Lett. B, 862: 139341 (2025)
	
	\bibitem{tumino2018}
	A. Tumino, C. Spitaleri, M. La Cognata, et al., Nature, 557 (7707): 687 (2018)
	
	\bibitem{li2016}
	K. A. Li, Y. H. Lam, C. Qi, et al., Phys. Rev. C, 94 (6): 065807 (2016)
	
	\bibitem{caughlan1988}
	G. R. Caughlan, W. A. Fowler, At. Data Nucl. Data Tables, 40 (2): 283-334 (1988)
	
\end{thebibliography}

\end{document}